\documentclass[preprint,showpacs,preprintnumbers,amsmath,amssymb,endfloa
ts*]{revtex4}
\usepackage{graphicx}

\newcommand{\bes}{\begin{eqnarray}}
\newcommand{\ees}{\end{eqnarray}}

\begin{document}

\thispagestyle{empty}
\title{
On the control of the lateral Casimir force between corrugated
surfaces
}
\author{E.~V.~Blagov,${}^{1}$ 
   G.~L.~Klimchitskaya,${}^{2}$
U.~Mohideen,${}^{3}$
 and
V.~M.~Mostepanenko${}^{1,2}$
}

\affiliation{${}^{1}$Noncommercial
Partnership ``Scientific Instruments'', Moscow, Russia
\\
${}^{2}$Departamento de F\'{\i}sica, Universidade Federal da
Para\'{\i}ba,
C.P.5008, CEP 58059--970, Jo\~{a}o Pessoa, Pb-Brazil \\
${}^{3}$Department of Physics, University of California,
Riverside, California 92521 
}

\begin{abstract}
The general perturbative expression for the lateral Casimir force
between two plates covered by longitudinal corrugations of
arbitrary shape is obtained. This expression is applicable for
corrugation periods larger than the separation distance. The cases 
of asymmetric corrugations are considered, which allow to increase
the maximum to minimum force ratio and affect the character of
equilibrium points. This opens new opportunities to control the
lateral Casimir forces for use in microelectromechanical
devices based entirely on the vacuum fluctuation properties. 
\end{abstract}

\pacs{12.20.Ds, 42.50.Lc}

\maketitle

It is well known that the Casimir force \cite{1} arises due to 
electromagnetic quantum fluctuations and acts between closely spaced 
surfaces. For short distances of order a nanometer, the limiting
form of the Casimir force results in the van der Waals force.
In fact the Casimir force is a collective quantum phenomenon which
results from the alteration of the zero-point photon spectrum
by surfaces of material bodies. H.~B.~G.~Casimir \cite{1} was
the first to calculate an extraordinary property that two parallel
uncharged metallic plates placed in vacuum at some separation $a$
would feel an attractive force per unit area 
$F_0(a)=-\pi^2\hbar c/(240a^4)$, where $\hbar$ is the Planck constant
and $c$ is the velocity of light. The Casimir force exhibits important
yet exotic dependencies on the shape of boundary surfaces, which are
not simple extensions from flat boundaries (see monographs \cite{2,3,4} 
and reviews \cite{5,6}).

Recently a number of experiments were performed on measuring
the normal Casimir force, acting perpendicular to the interacting
surfaces \cite{7,8,9,10,11,12,13,14,15}.
The highest precision of order of 1\% of the measured force was
achieved in the measurements by means of atomic
force microscope \cite{8,9,10,11} and microelectromechanical
torsional oscillator \cite{14,15}. The extent of agreement
between experiment and theory was used to set stringent limits
on predictions of fundamental physical theories \cite{15,16,17,18,19}.
The Casimir force can be the dominant force acting between surfaces for
separation distances on the order or less than a few hundred
nanometers. The Casimir effect was used to produce mechanical
motion of microfabricated silicon plate \cite{20}. This is the first 
case of microelectromechanical
device which shows actuation by the Casimir force. Note also that
the Casimir forces acting between closely spaced surfaces play
the crucial role in the fabrication and yield of
microdevices due to
the phenomena of adhesion and stiction \cite{21}.

Similar to the normal Casimir force, the lateral Casimir force 
originates from the modifications of zero-point 
oscillations by material boundaries. It acts tangential to the
two surfaces with nanoscale periodic corrugations. To first order
the origin of the lateral Casimir force can be simply understood 
as follows. As the Casimir force is a strong function of distance
($F_0\sim a^{-4}$), the natural tendency is for corrugated plates
to align themselves such that peaks of the top plate are directly
over the peaks of the bottom plate. This force to align the peaks is
then the origin of the lateral force acting in the horizontal direction
between the plates. The lateral Casimir force for
anisotropic boundaries was investigated theoretically and a harmonic
dependence on corresponding angle was predicted \cite{3,22,23}.
For two aligned sinusoidally corrugated plates made of ideal metal
the lateral Casimir force was discussed in Refs.~\cite{24,25} and
a harmonic dependence on a phase shift between 
corrugations was found. In Ref.~\cite{26} the first experimental
observation of the lateral Casimir force between the sinusoidally
corrugated plate and large corrugated sphere was reported.
The predicted sinusoidal dependence on the phase shift between
corrugations was confirmed. The analysis of the theoretical
dependence for the lateral Casimir force was performed in 
Ref.~\cite{26} and the optimum values of the parameters
providing the maximum values of the lateral force in the case
of sinusoidal corrugations were found.

In this report we investigate the lateral Casimir force between
two parallel plates covered by longitudinal grooves of arbitrary
shape. This opens new opportunities to change the magnitude of
the lateral Casimir force and obtain asymmetric lateral
forces with a more complicated character of equilibrium points
leading to a rich variety of surface interactions.
As a consequence, it may become possible to control friction
through nanoscale modification of the contact surfaces
to increase the yield of microdevices \cite{21}.

The normal Casimir force acting between plates 
covered with small distortions has been the subject of much attention 
in literature (see Ref.~\cite{6} for review). In Ref.~\cite{27}
a general method was provided which enables one to find corrections
to the normal Casimir force in the configuration of two parallel plates
with small deviations from plane parallel geometry. This method is based
on the pairwise summation of the interatomic Casimir potentials with
a subsequent normalization of the obtained interaction constant
\cite{3,6}. In doing so, the amplitudes of deviations
{}from plane parallel geometry $A_{1,2}$ are assumed to be small 
as compared to the separation $a$ between plates.

To fix the definition of the separation between the nonplanar
plates, we describe their surfaces by the equations
\begin{equation}
z_1^{(s)}=A_1f_1(x,y), \qquad
z_2^{(s)}=a+A_2f_2(x,y),
\label{eq1}
\end{equation}
\noindent
where $a$ is the mean value of the distance between the square plates
with a side $2L$ and the coordinate system is connected with the lower
plate. The values of the amplitudes are chosen in such a way that
$\max|f_i(x,y)|=1$. The zero point in the $z$-axis is so chosen that
\begin{eqnarray}
&&
\langle z_1^{(s)}\rangle\equiv A_1\langle f_1(x,y)\rangle
=\frac{A_1}{4L^2}\int_{-L}^{L}dx\int_{-L}^{L}dyf_1(x,y)=0,
\nonumber \\
&&
\langle z_2^{(s)}\rangle\equiv a+A_2\langle f_2(x,y)\rangle=a.
\label{eq2}
\end{eqnarray}

The normal Casimir force between the plates covered with distortions
was represented by a perturbative expansion up to the fourth order
\cite{27}
\begin{equation}
F(a)=F_0(a)\sum\limits_{k=0}^{4}\sum\limits_{l=0}^{4-k}
c_{kl}\left(\frac{A_1}{a}\right)^k\left(\frac{A_2}{a}\right)^l,
\label{eq3}
\end{equation}
\noindent
where $F_0(a)$ is the Casimir force per unit area of flat plates
defined earlier, $c_{00}=1$ and $c_{01}=c_{10}=0$
[this follows from our choice (\ref{eq2})]. Explicit expressions for
the other coefficients $c_{kl}$ are found in Ref.~\cite{27}
(see also Refs.~\cite{3,6}). They are rather combersome. If, however, 
both functions $f_1$ and $f_2$ are periodic in two variables
with periods much larger than $a$, but much less than $2L$, 
simple approximate expressions for $c_{kl}$ follow leading to
\begin{eqnarray}
&&F(a)=F_0(a)\left[
\vphantom{\left(\langle f_1^2\rangle A_1^2\right)}\right.
1+
\frac{10}{a^2}\left(
\langle f_1^2\rangle A_1^2-2\langle f_1f_2\rangle A_1A_2
+\langle f_2^2\rangle A_2^2\right)
\nonumber \\
&&\phantom{aaa}
+\frac{20}{a^3}\left(
\langle f_1^3\rangle A_1^3-3\langle f_1^2f_2\rangle A_1^2A_2
+3\langle f_1f_2^2\rangle A_1A_2^2
-\langle f_2^3\rangle A_2^3\right)
\label{eq4} \\
&&\phantom{aaa}
+\frac{35}{a^4}\left.\left(
\langle f_1^4\rangle A_1^4-4\langle f_1^3f_2\rangle A_1^3A_2
+6\langle f_1^2f_2^2\rangle A_1^2A_2^2-
4\langle f_1f_2^3\rangle A_1A_2^3
+\langle f_2^4\rangle A_2^4\right)\right].
\nonumber
\end{eqnarray}
\noindent
Here the averaging as in Eq.~(\ref{eq2}) is done over the periods.

{}From Eq.~(\ref{eq4}) the 
Casimir energy per unit area can be found
\begin{eqnarray}
&&E(a)=\int_{a}^{\infty}F(a^{\prime})da^{\prime}=
E_0(a)\left[
\vphantom{\left(\langle f_1^2\rangle A_1^2\right)}\right.
1+
\frac{6}{a^2}\left(
\langle f_1^2\rangle A_1^2-2\langle f_1f_2\rangle A_1A_2
+\langle f_2^2\rangle A_2^2\right)
\nonumber \\
&&\phantom{aaa}
+\frac{10}{a^3}\left(
\langle f_1^3\rangle A_1^3-3\langle f_1^2f_2\rangle A_1^2A_2
+3\langle f_1f_2^2\rangle A_1A_2^2
-\langle f_2^3\rangle A_2^3\right)
\label{eq5} \\
&&\phantom{aaa}
+\frac{15}{a^4}\left.\left(
\langle f_1^4\rangle A_1^4-4\langle f_1^3f_2\rangle A_1^3A_2
+6\langle f_1^2f_2^2\rangle A_1^2A_2^2-
4\langle f_1f_2^3\rangle A_1A_2^3
+\langle f_2^4\rangle A_2^4\right)\right],
\nonumber
\end{eqnarray}
\noindent
where $E_0(a)=-\pi^2\hbar c/(720a^3)$ is the Casimir energy 
between two flat plates.

Now let functions $f_{1,2}$ depend on only one variable,
say $x$, i.e. they describe longitudinal corrugations extending in the
 $y$ direction. Let their periods be equal, 
$\Lambda_1=\Lambda_2=\Lambda$, but there may exist  some phase
shift $x_0$.
In this case $E(a)=E(a,x_0)$ and the lateral Casimir force arises
directed along $x$-axis. It can be found from Eq.~(\ref{eq5}) as
\begin{eqnarray}
&&F^{lat}(a,x_0)=-\frac{\partial E(a,x_0)}{\partial x_0}=
F_0(a)\frac{2A_1A_2}{a^2}\left[
2\frac{\partial}{\partial x_0}\langle f_1f_2\rangle 
\right.
\nonumber \\
&&\phantom{aaa}
+5\left(
\frac{A_1}{a}\frac{\partial}{\partial x_0}\langle f_1^2f_2\rangle
-\frac{A_2}{a}\frac{\partial}{\partial x_0}\langle f_1f_2^2\rangle
\right)
\label{eq6} \\
&&\phantom{aaa}
+10\left.\left(
\frac{A_1^2}{a^2}\frac{\partial}{\partial x_0}\langle f_1^3f_2\rangle 
-\frac{3}{2}\frac{A_1A_2}{a^2}\frac{\partial}{\partial x_0}
\langle f_1^2f_2^2\rangle
+\frac{A_2^2}{a^2}\frac{\partial}{\partial x_0}
\langle f_1f_2^3\rangle\right)\right].
\nonumber
\end{eqnarray}

Eq.~(\ref{eq6}) permits to find the lateral Casimir
force between plates covered by the longitudinal corrugations of any 
shape. Under the condition used  $\Lambda\gg a$ the precision of the 
pairwise summation method is very high and the results practically
coincide with the exact ones (this was demonstrated in Ref.~\cite{25}
for the case of sinusoidal corrugations). The same
conclusion can be obtained by the method of Ref.~\cite{28}.
The advantage of Eq.~(\ref{eq6}) is the possibility to quickly
obtain the results for corrugations of any shape and to generalize
this formalism for the case of real metals of finite conductivity
(see Ref.~\cite{26}).

As was shown in Ref.~\cite{26} for the sinusoidal 
corrugations, the dependence of the lateral force on the phase
shift is almost sinusoidal.
Specifically, the points of the stable and unstable equilibrium
(where $F^{lat}=0$) alternate with half the corrugation period. 
The magnitudes of the maximum and minimum values of $F^{lat}$
are equal to each other so that their ratio is equal to unity.
These characteristic features of the lateral Casimir force
are preserved for any longitudinal corrugation whose shape is
symmetric relatively to some vertical axis.

In applications to microelectromechanical systems, a more complicated
nature of the lateral Casimir force may be desirable, i.e. different
magnitudes of maximum and minimum values and a more complicated
character
of the points of equilibrium. This can be achieved by use of 
asymmetric longitudinal grooves on the plate surfaces. We consider
saw toothed corrugations with equal amplitudes shown in Fig.~1,a.
Within the periods (from 0 to $\Lambda$ for $f_1$ and from $x_0$ to
$x_0+\Lambda$ for $f_2$) the analytical representations of the
corrugation functions of Fig.~1,a are
\begin{equation}
f_1(x)=\frac{2x}{\Lambda}-1, \qquad
f_2(x)=1-\frac{2(x-x_0)}{\Lambda}.
\label{eq7}
\end{equation}
\noindent
Calculating all matrix elements from Eq.~(\ref{eq6}) over the
period from 0 to $\Lambda$, one obtains
\begin{eqnarray}
&&
\langle f_1f_2\rangle=-\frac{1}{3}+2\frac{x_0}{\Lambda}-
2\frac{x_0^2}{\Lambda^2},
\nonumber \\
&&
\langle f_1^2f_2\rangle=
\langle f_1f_2^2\rangle=-\frac{4}{3}\frac{x_0}{\Lambda}+
4\frac{x_0^2}{\Lambda^2}-
\frac{8}{3}\frac{x_0^3}{\Lambda^3},
\label{eq8} \\
&&
\langle f_1^3f_2\rangle=
\langle f_1f_2^3\rangle=-\frac{1}{5}+2\frac{x_0}{\Lambda}-
6\frac{x_0^2}{\Lambda^2}+
8\frac{x_0^3}{\Lambda^3}-4\frac{x_0^4}{\Lambda^4},
\nonumber \\
&&
\langle f_1^2f_2^2\rangle=\frac{1}{5}
-\frac{8}{3}\frac{x_0^2}{\Lambda^2}+
\frac{16}{3}\frac{x_0^3}{\Lambda^3}-
\frac{8}{3}\frac{x_0^4}{\Lambda^4}.
\nonumber
\end{eqnarray}

Substitution of Eq.~(\ref{eq8}) into Eq.~(\ref{eq6}) with
$A_1=A_2=A$ leads to the result
\begin{equation}
F^{lat}(a,x_0)=8|F_0(a)|\frac{A^2}{a\Lambda}
\left(2\frac{x_0}{\Lambda}-1\right)
\left[1+10\frac{A^2}{a^2}\left(1-2\frac{x_0}{\Lambda}
+2\frac{x_0^2}{\Lambda^2}\right)\right].
\label{eq9}
\end{equation}

As an illustration, Fig.~2,a shows the dependence of
$F^{lat}/|F_0|$ on $x_0/\Lambda$ computed by Eq.~(\ref{eq9}) with
the typical values of parameters $A/a=0.3$ and $a/\Lambda=0.2$.
{}From this figure it is seen that the lateral force is still
symmetric and the points of unstable equilibrium 
($x_0/\Lambda=0.5,\,1.5,\,\ldots$) are exactly in the middle
between the points of stable equilibrium 
($x_0/\Lambda=0,\,1,\,2,\,\ldots$). Like in the case of the
sinusoidal corrugations, the lateral force is negative over one
half of the period and positive over another half. The magnitudes
of maximum and minimum values of the force are equal. At the
same time, the case of saw toothed structures is
different from sinusoidal corrugations because here the extremum
values of the lateral force are achieved near the points of
stable equilibrium (where the force is discontinuous).
The points of stable equilibrium in this configuration are
especially stable. Even small deviation {}from the
stable equilibrium (where the value of the lateral force is taken 
equal to zero, i.e., half a sum of the limiting values {}from the left
and {}from the right) leads to a large
lateral force, restoring the state of equilibrium.
Thus, the plates with saw toothed corrugations could be used
in microdevices where the lateral
displacements of the elements should be avoided (e.g. in devices such
as micromirrors, microgears, micropumps, microsensors and microvalves).

Now we consider even more asymmetric longitudinal corrugations
on the lower plate allowing to obtain different magnitudes for
the maximum and minimum values of the lateral Casimir force.
On the upper plate the same corrugations as in the previous
example are preserved. The profiles are shown in Fig.~1,b.
The new parameter $\Delta=l_x/\Lambda$ characterizes the
extent of asymmetry. If $l_x=0$ ($\Delta=0$), the profiles in
Fig.~1,b coincide with those in Fig.~1,a. The function $f_1$,
describing the lower plate, can be presented as (within one
period)
\begin{equation}
f_1(x)=\left\{
{\begin{array}{rr}
-\frac{1-\Delta}{1+\Delta}, & 0<x\leq l_x, \\
\frac{2}{1-\Delta^2}\frac{x}{\Lambda}-
\frac{1+\Delta^2}{1-\Delta^2},& l_x<x\leq\Lambda.
\end{array}}
\right.
\label{eq10}
\end{equation}
\noindent
The function $f_2$ for the upper plate is given by Eq.~(\ref{eq7}).
Note that according to Eq.~(\ref{eq2}) both functions have
zero mean values over the period.

The expression for the lateral Casimir force is
obtained by calculating the matrix elements from Eq.~(\ref{eq6}) over
the period 
(0,$\Lambda$). Here the result takes a different form
depending on
whether $x_0\leq l_x$ or $x_0\geq l_x$. As before, we consider $A_1=A_2$. 
Then for $x_0\leq l_x$ one obtains

\begin{eqnarray}
&&
F^{lat}(a,x_0)=-8|f_0(a)|\,\frac{A^2}{a\lambda}\,
\frac{1-\Delta}{1+\Delta}
\left\{
\vphantom{\left[\frac{1+5\Delta^2+4\Delta^3+
\Delta^4}{(1+\Delta)^2}\right]}
1+\frac{10}{3}\frac{A}{a}\frac{1}{1+\Delta}
\left[\Delta(3+\Delta)-3\frac{x_0}{\Lambda}(1+\Delta)\right]\right.
\nonumber \\
&&\phantom{aaa}
\left.
+10\frac{A^2}{a^2}\left[\frac{1+5\Delta^2+4\Delta^3+
\Delta^4}{(1+\Delta)^2}-4\frac{x_0}{\Lambda}
\frac{\Delta(3+\Delta)}{1+\Delta}+6\frac{x_0^2}{\Lambda^2}
\right]\right\}.
\label{eq11}
\end{eqnarray}
\noindent
For large phase shifts $x_0\geq l_x$ the expression for the lateral
Casimir force is a bit more complicated
\begin{equation}
F^{lat}(a,x_0)=8|f_0(a)|\frac{A^2}{a\lambda}\frac{1}{1-\Delta^2}
\left(2\frac{x_0}{\Lambda}-1-\Delta^2\right)
\left(1+\frac{10}{3}\frac{A}{a}X_1+ 
10\frac{A^2}{a^2}X_2\right),
\label{eq12}
\end{equation}
\noindent
where the coefficients $X_{1,2}$ are given by
\begin{eqnarray}
&&
X_1=-\frac{\Delta^2\left[2-3\Delta+3\Delta^2+\Delta^3-
3\left(1+\Delta^2\right)x_0/\Lambda+
3x_0^2/\Lambda^2\right]}{\left(1-\Delta^2\right)\left(1+
\Delta^2-2x_0/\Lambda\right)},
\label{eq13} \\
&&
X_2=\frac{1}{\left(1-\Delta^2\right)^2\left(1+
\Delta^2-2x_0/\Lambda\right)}\left[
\vphantom{\frac{x_0^2}{\Lambda^2}\left(1+\Delta^6\right)}
1-\Delta^2+10\Delta^4-12\Delta^5+\Delta^6\right.
\nonumber \\
&&
\phantom{aaa}
+4\Delta^7+\Delta^8-4\frac{x_0}{\Lambda}\left(
1-\Delta^2+3\Delta^3-4\Delta^5+3\Delta^6+\Delta^7\right)
\nonumber \\
&&
\phantom{aaa}
\left.
+6\frac{x_0^2}{\Lambda^2}\left(1+\Delta^6\right)
-4\frac{x_0^3}{\Lambda^3}\left(1-\Delta^2+\Delta^4\right)
\right].
\nonumber
\end{eqnarray}
\noindent
Direct calculation shows that for $x_0=l_x$ Eqs.~(\ref{eq11})
and (\ref{eq12}) lead to one and the same result. Furthermore, for
$\Delta=0$ one obtains $X_1=0$, 
$X_2=1-2(x_0/\Lambda)+2(x_0/\Lambda)^2$, so that Eq.~(\ref{eq12})
coincides with Eq.~(\ref{eq9}) as required (because the
profiles in Figs.~1,a,b coincide when
$\Delta=0$).

The coordinate of the point of unstable equilibrium, where 
$F^{lat}(a,{\tilde{x}}_0)=0$, is found from Eq.~(\ref{eq12})
\begin{equation}
{\tilde{x}}_0=\frac{\Lambda\left(1+\Delta^2\right)}{2}.
\label{eq14}
\end{equation}
\noindent
For example, in Fig.~2,b one period of the relative lateral
Casimir force $F^{lat}/|F_0|$ is plotted as a function of
$x_0/\Lambda$ for $\Delta=1/2$ ($l_x=0.5\Lambda$),
$A/a=0.3$ and $a/\Lambda=0.2$.
According to Eq.~(\ref{eq14}), the point of unstable equilibrium
is ${\tilde{x}}_0=5\Lambda/8$, i.e. it is asymmetric being shifted from
the middle
of a period. It is seen also that the magnitudes of maximum and
minimum values of the lateral Casimir force in this case differ
by 1.9. Needless to say, the work done by the lateral
force over one period is equal to zero as it should be in any 
adiabatic process. If $\Delta$ is further increased, the position
of the points of unstable equilibrium will be even more shifted
to the right boundary of the period with an increase of the ratio
of magnitudes of the maximum to minimum lateral forces.

In the above the method of pairwise summation and perturbation
theory in relative distortion amplitudes were applied to obtain
the general expression for the lateral Casimir force between
metallic plates with longitudinal corrugations of arbitrary
profile. The pairwise summation works well when the corrugation
period is several times larger than the separation between
plates. The obtained expressions were used in the case of
asymmetric saw tooth like structures which lack the right-left
symmetry of sinusoidal corrugations. It was shown that with the
proper choice of the corrugation shape and parameters, it is
possible to change not only the magnitude of the lateral Casimir 
force, but make it asymmetric and affect the character of points
of equilibrium. In particular, the maximum to
minimum force ratio can be increased by several times.
This opens new opportunities to control the lateral Casimir force 
for the diversified applications in microelectromechanical
devices based entirely on the vacuum fluctuation properties
of quantum electrodynamics.

This work was supported
by the National Institute of Standards and Technology, through a
Precision Measurement Grant and the University of California and
Los Alamos National Laboratory through the LANL-CARE program. G.L.K. and
V.M.M. were also partially supported by CNPq and
Finep (Brazil).

\begin{figure*}
\vspace*{-2cm}
\includegraphics{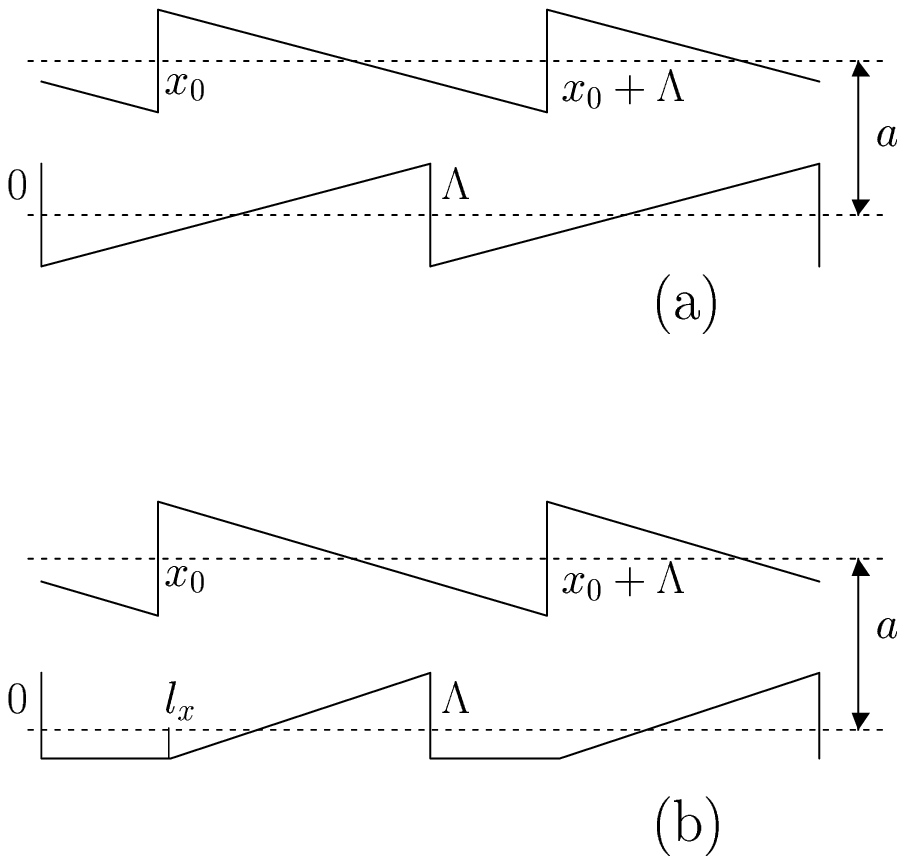}
\vspace*{-12cm}
\caption{
The saw tooth corrugations on both plates with a phase shift
$x_0$ and period $\Lambda$ (a).
On lower plate the flat segments of length $l_x$ are
added, while preserving the same corrugations on upper plate (b).
}
\end{figure*}
\begin{figure}
\vspace*{0cm}
\includegraphics{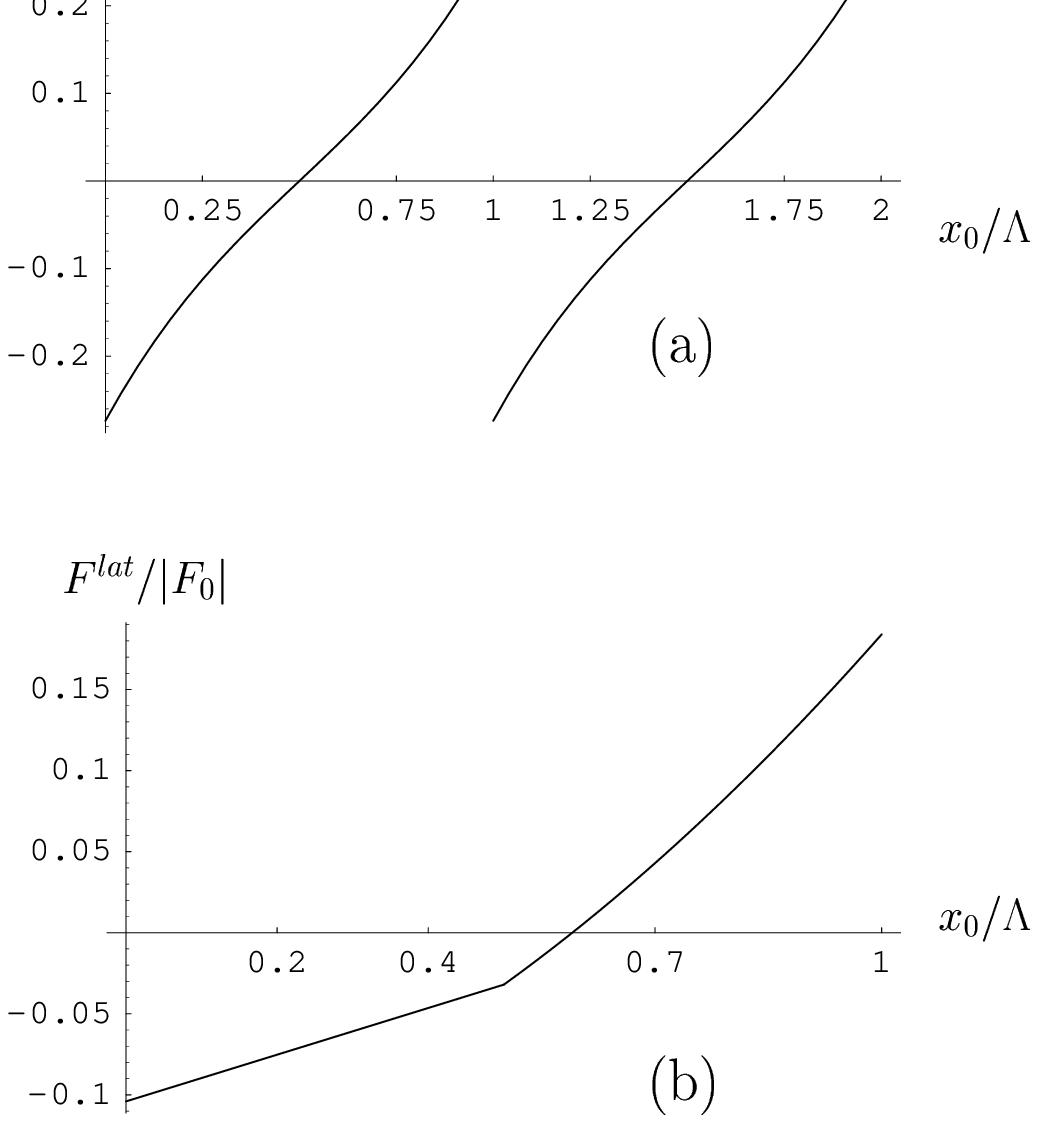}
\vspace*{-10cm}
\caption{
The relative lateral Casimir force as a function of a relative 
phase shift for the corrugations of Fig.~1,a and Fig.~1,b
[(a), (b), respectively].
}
\end{figure}
\end{document}